\documentclass[a4paper, amsfonts, amssymb, amsmath, reprint, showkeys, nofootinbib, twoside,superscriptaddress]{revtex4-1}
\usepackage[english]{babel}
\usepackage[utf8]{inputenc}
\usepackage{amsthm}
\usepackage{mathtools}
\usepackage{xcolor}
\usepackage{graphicx}
\usepackage[left=23mm,right=13mm,top=35mm,columnsep=15pt]{geometry} 
\usepackage{adjustbox}
\usepackage{placeins}
\usepackage[T1]{fontenc}
\usepackage{lipsum}
\usepackage{csquotes}
\usepackage[colorinlistoftodos, color=green!40, prependcaption]{todonotes}
\usepackage{tcolorbox}
\usepackage{graphicx}
\usepackage{caption}
\usepackage{subcaption}
\usepackage{subeqnarray}

\usepackage[pdftex, pdftitle={Article}, pdfauthor={Author}]{hyperref} 
\usepackage{xcolor}

\usepackage{graphicx}

\newcommand{\bs}{\boldsymbol}

\begin{document}
\title{Paragliders' Launch Trajectory is Universal}


\author{Quentin Da Cruz Lopes}
\affiliation{LadHyX UMR CNRS 7646, \'Ecole polytechnique, 91128 Palaiseau Cedex, France\medskip}

\author{Sophie Ramananarivo}
\affiliation{LadHyX UMR CNRS 7646, \'Ecole polytechnique, 91128 Palaiseau Cedex, France\medskip}

\author{Caroline Cohen }
\affiliation{LadHyX UMR CNRS 7646, \'Ecole polytechnique, 91128 Palaiseau Cedex, France\medskip}

\author{Michael Benzaquen}
\email{michael.benzaquen@polytechnique.edu}
\affiliation{LadHyX UMR CNRS 7646, \'Ecole polytechnique, 91128 Palaiseau Cedex, France\medskip}

\date{\today} 

\begin{abstract}
We designed and built a reduced-scale model experiment to study the paragliding inflation and launching phase at given traction force. We show that the launch trajectory of a {single skin} glider is universal, that is, independent of the exerted force. As a consequence,  the length of the take-off run required for the glider to reach its “ready to launch” vertical position is also universal. We successfully compare our results to full-scale experiments, and show that such universality can be understood through a simple theoretical model.
\end{abstract}


\maketitle

\section*{Introduction} \label{sec:intro}

Paragliding is a young adventure sport dating back to the early 1980s, in which a lightweight wing with no rigid primary structure is launched by a running pilot.
 For a physicist, it truly is a endless source of fascinating unexplored problems, combining a variety of fields from fluid mechanics to fluid-structure interactions, and including flight mechanics, materials science, micrometeorology, and even game theory in the context of understanding exploration-exploitation  strategies of thermal convection~\cite{landell2022paragliding,manuel}. {Indeed, cross-country flying combines climbs using thermal updrafts and glides through stationary air. Race pilots seeking to cover the greatest possible distance must find the right balance between  the individualistic strategy based on one’s knowledge to find the next nascent thermal, with the risk of missing out, and  the collective strategy consisting in following other pilots who concentrate where the thermals are located, with the risk of getting there too late.}

Since the first prototypes, paraglider wings have continuously evolved, both in terms of performance and security. Most of the research done by paragliding manufacturers  has focused on optimising wings for steady flight phases~\cite{falquier2019longitudinal,benedetti2012paragliders}, and unsteady regimes have only received limited attention~\cite{muller2018modelling}. 
In particular, many questions remain unsolved when it comes to the dynamics of stalls or wing collapses. Further  research may therefore provide quantitative elements to improve the safety of modern gliders, the airworthiness certification of which is now based on the rather qualitative feelings of test pilots. 
In addition, accidentology studies show that a substantial fraction of accidents occur during take-off~\cite{soleil2016accidentologie,fasching1997paragliding}, which makes  the study of the launch phase crucial. 

\begin{figure}[b!]
\includegraphics[width=0.9\columnwidth]{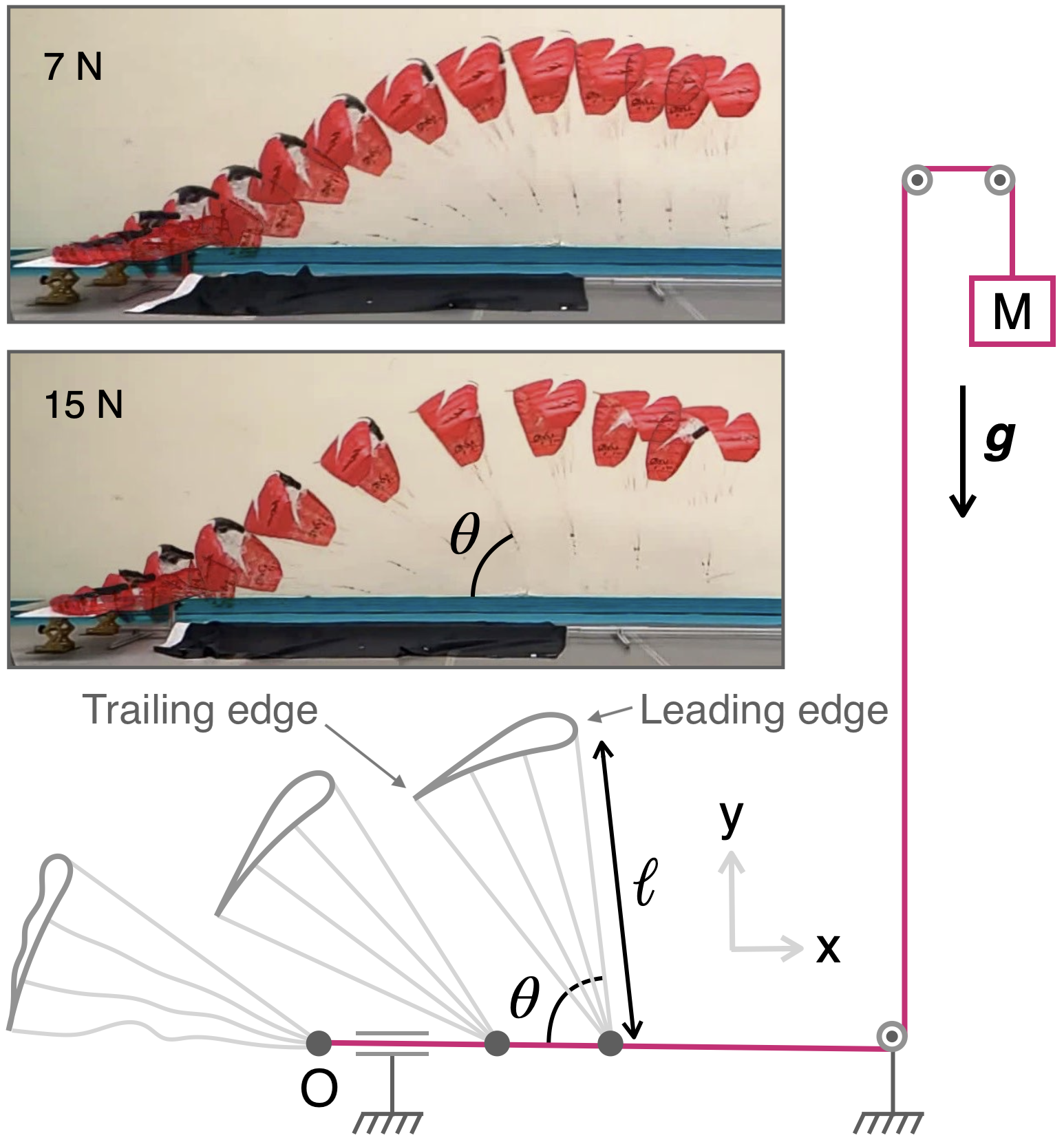}
\captionsetup{justification=raggedright,
singlelinecheck=false
}
    \caption{(Color online) Reduced-scale experiment (see Sec.~\ref{sec:redexp}). The top left captions show chronophotographies of the experiment with  $Mg=7.0\,$N and $15.0\,$N respectively. The timestep between shots is 40\,ms. The overall dimensions of the setup are $4\,$m along $x$ and $y$, and 0.5$\,$m along $z$. }
    \label{fig:redexp}
\end{figure}

In this paper, we investigate the dynamics of the launching phase {of a single skin wing}: How does a seemingly simple piece of fabric inflate when pulled by the pilot to become a reasonably stable aircraft in just a matter of seconds? To answer this question, we designed and built a reduced-scale model experiment and studied the paragliding inflation and launching phases at  given traction forces. We found that the launch trajectory is universal -- in the sense that it does not depend on the strength of the exerted force -- and, as a result, neither does the distance required for the glider to reach its “ready to launch” vertical position. Due to the  limitations of the model experiment, namely the difficulty of scaling down the stiffness of the materials (fabric, lines linking the fabric to the risers which, in turn, are connected to the harness \cite{jargon}), we also performed full-scale experiments on the field that showed excellent agreement with the reduced-scale experimental results.

\section{Reduced-scale experiment}\label{sec:redexp}

\begin{figure}[t!]
\includegraphics[width=0.9\columnwidth]{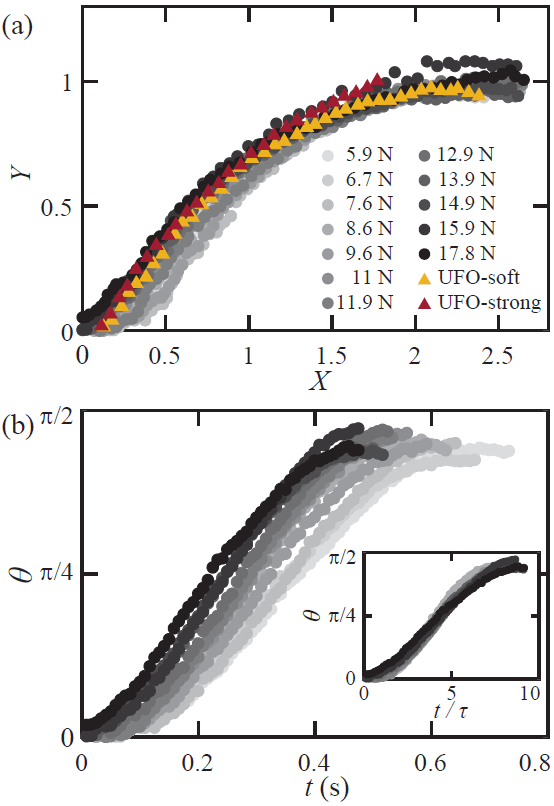}
\captionsetup{justification=raggedright,
singlelinecheck=false
}
    \caption{(Color online) (a) Dimensionless trajectory of the trailing edge of a reduced-scale glider during launching at constant traction force with $X=x/\ell$ and $Y=y/\ell$, for different values of the traction force (denoted by  grayscale dots). Dimensionless trajectories for a real-scale {AirDesign}$^{\text{\textregistered}}$ UFO glider under low and strong traction force are shown with triangles. 
    (b)~Time evolution of the glider's angle $\theta$. Rescaling of time with the characteristic timescale $\tau=\sqrt{m\ell/F_O}$ collapses the data (inset). }
    \label{fig:masterplot}
\end{figure}

In order to work in a controlled environment to ensure reproducibility, we started with a reduced-scale experiment in which a small {single-skin} paraglider  (Oxy 0.5 model designed by \textit{Opale Paramodels}, see characteristics in Tab.~\ref{table_car}) is pulled from its risers along a horizontal 3\,m-long guide rail by a wire and pulley system (Fig.~\ref{fig:redexp}). To minimize friction, the paraglider is fastened to a ring looped around a string that serves as the guide.

For the experiment to be as realistic as possible, we chose to work with a specified traction force rather than a specified velocity.  This is because, when pilots launch, they aim to exert a nearly constant physical effort in order to achieve good control of the wing. While such effort is most certainly not perfectly constant over the launching phase, assuming constant force is a good approximation.   
Here, the constant force is provided by a falling mass $M$ attached to the wire.

 During the launching phase, we identify two  processes: (i) the chambers fill with air to give the aircraft its wing shape, thus allowing lift, and (ii) the ``inflated'' wing rises from the horizontal to the vertical position. These two processes overlap somewhat in time, but we analyze them as separate phases. Our choice of a single-skin glider results from the fact that the time overlap between phases (i) and (ii) is observed to be much weaker compared to a regular double-skin glider, due to the much faster dynamics of phase (i). In the top left panels of Fig.~\ref{fig:redexp}, one can see that the glider is fully inflated when the trailing edge used to track its position  leaves the ground (this defines $t=0$ in the following). See Sec.~\ref{sec:conclusion} for a discussion on regular double-surface gliders.

\begin{table}[h!]
        \centering
        \begin{tabular}{|l|l|l|}
            \hline
            Glider & Oxy 0.5 &   UFO 13 \\
            \hline
           Flat wingspan (m) & 1.3 & 8.04 \\
            \hline
            Flat area (m$^2$) &0.5  &13.0  \\
            \hline
           Aspect ratio& 4.2&4.9 \\
            \hline                     
          {Average line length $\ell$  (m)$^*$ } &{0.98} &{6.05} \\
            \hline
            Chambers  & 17 & 27 \\
            \hline
             Weight (kg)  & 0.05 & 1.36 \\
            \hline
         
        \end{tabular}
        \captionsetup{justification=raggedright,
singlelinecheck=false
}
        \caption{Glider characteristics for the Oxy 0.5 wing (\textit{Opale Paramodels}) and the   UFO 13 (\textit{AirDesign Gliders}). $^*$For the detailed line charts see manufacturer's websites~\cite{AD,Opale}.}\label{table_car}
    \end{table}

    \begin{figure*}[t!]
\captionsetup{justification=raggedright,
singlelinecheck=false
}
\centering
\includegraphics[width=\textwidth]{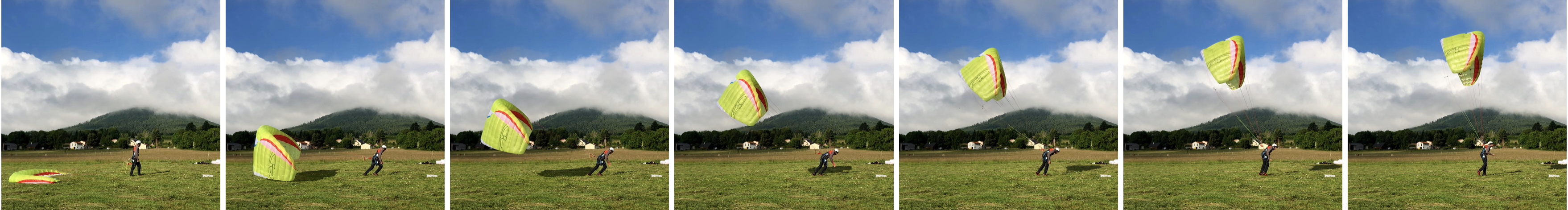}
\caption{(Color online) Pictures of the real scale experiment. The timestep between shots is 200\,ms.}
    \label{fig:alex}
\end{figure*}

Figure~\ref{fig:masterplot}a displays the dimensionless trajectory $Y(X)$ of the trailing edge during launching runs with different values of the traction force $F_O$, ranging from $5.9$ to $17.8$\,N. The trailing edge position is tracked manually using  ImageJ~\cite{imageJ} for image analysis. Note that the  string that is used as a guide rail tends to distort a little under the pulling action of the glider. To account for this upwards shift, we subtract the height of the attachment point from that of the trailing edge. Strikingly, all the dimensionless trajectories fall on top of each other, thereby indicating that the dimensionless launch trajectory is universal and does not depend on the traction force $F_O$. However, the traction force does determine the speed at which the paraglider moves along this universal trajectory, as shown in the time-evolution of its angle $\theta(t)$ in Fig.~\ref{fig:masterplot}b, where $\theta$ is the angle of the risers with respect to the horizontal (Fig.~1). As shall be discussed in Section~\ref{sec:theory},  $F_O$ sets the characteristic time scale of the ascent $\tau=\sqrt{m\ell/F_O}$ of a wing of mass $m$: the glider's angle data all collapse on a single curve when plotted as a function of $t/\tau$ (inset of Fig.~\ref{fig:masterplot}b).

\section{Full-scale experiment}\label{sec:realexp}

Due to our concerns discussed above about downscaling all the characteristics of s single-skin glider, we decided to compare the above results to full-scale experiments (see Fig.~\ref{fig:alex}). Using a 13\,m$^2$ flat single-surface glider (AirDesign$^{\text{\textregistered}}$ gliders UFO wing, see characteristics in Tab.~\ref{table_car}), we were able to record a few launches on a calm day at Puy de Dôme (Auvergne, France). A seasoned test pilot was asked to launch with low and then with strong traction strength respectively. Unlike in laboratory experiments, there was a slight wind  of about $7$ m{$\cdot$}s$^{-1}$, which we removed when computing the horizontal velocity of the wing and plotting the trajectories in Fig.~\ref{fig:masterplot}a (in triangles). As one can see, the trajectories collapse with their reduced-scale counterparts, thereby validating the reduced-scale methodology.

{Note that, in addition to the non-scalability of the stiffness of the materials, the Oxy wing is not a perfect homothetic transformation of the UFO glider. While they are both single skin wings with similar geometric features, differences exist (see Table~\ref{table_car}) that may lead to different aerodynamic properties.
 It would be interesting to conduct more experiments on different gliders with a wider range of geometric characteristics. This is, however, beyond the scope of the present paper.}

\section{Theoretical model}
\label{sec:theory}

In this section, we present a simple theoretical model to account for the experimental results. We assume that a rigid glider of mass $m$ is characterized by its lift and drag coefficients~\cite{ross1993computational,moreno2017aerodynamic}, commonly denoted $C_L(\alpha)$ and $C_D(\alpha)$ respectively, 
 where $\alpha$ is the angle of attack (Fig.~4). The lift and drag forces are applied to the point $M$ that is called the center of pressure or the aerodynamic center.  They take the form:
\begin{subeqnarray}
\bs L =\textstyle \frac12 \rho S C_L(\alpha)\bs V_M^2 \bs e_\perp \slabel{eq:lift}\\
\bs D =\textstyle \frac12 \rho S C_D(\alpha)\bs V_M^2 \bs e_\parallel,\slabel{eq:drag}
\end{subeqnarray}
with $\rho$ the air density, $S$ the projected wing area, $V_M$ the magnitude of $M$'s velocity in the lab's frame of reference (or true airspeed). $\boldsymbol e_\parallel$ and $\boldsymbol e_\perp$ are the unit vectors respectively parallel and orthogonal to the air flow in the frame of reference of the wing. Note that the projected area (area of the fully inflated wing's shadow on the ground when the sun is at the zenith) is slightly smaller than the flat area (total area of wing fabric when laid on the ground) because of the profile curvature. 
We further assume that the center of pressure is attached to the traction point $O$ with a rigid weightless line of length $\ell$ exerting a tension $\pm \bs T_e = \mp T_e \bs e_r$ on $M$ and $O$ respectively, with $\boldsymbol e_r=\boldsymbol{OM}/OM$. Finally, $O$ is constrained to move along the $x$ axis only (equivalently, $O$ is subject to a vertical force $\boldsymbol R$, with $R=T_e \sin \theta$, see Fig.~4), and is pulled by a horizontal force $\boldsymbol F_O = F_O \bs e_x$.

 \begin{figure}[b!]
\captionsetup{justification=raggedright,
singlelinecheck=false
}
\includegraphics[width=0.9\columnwidth]{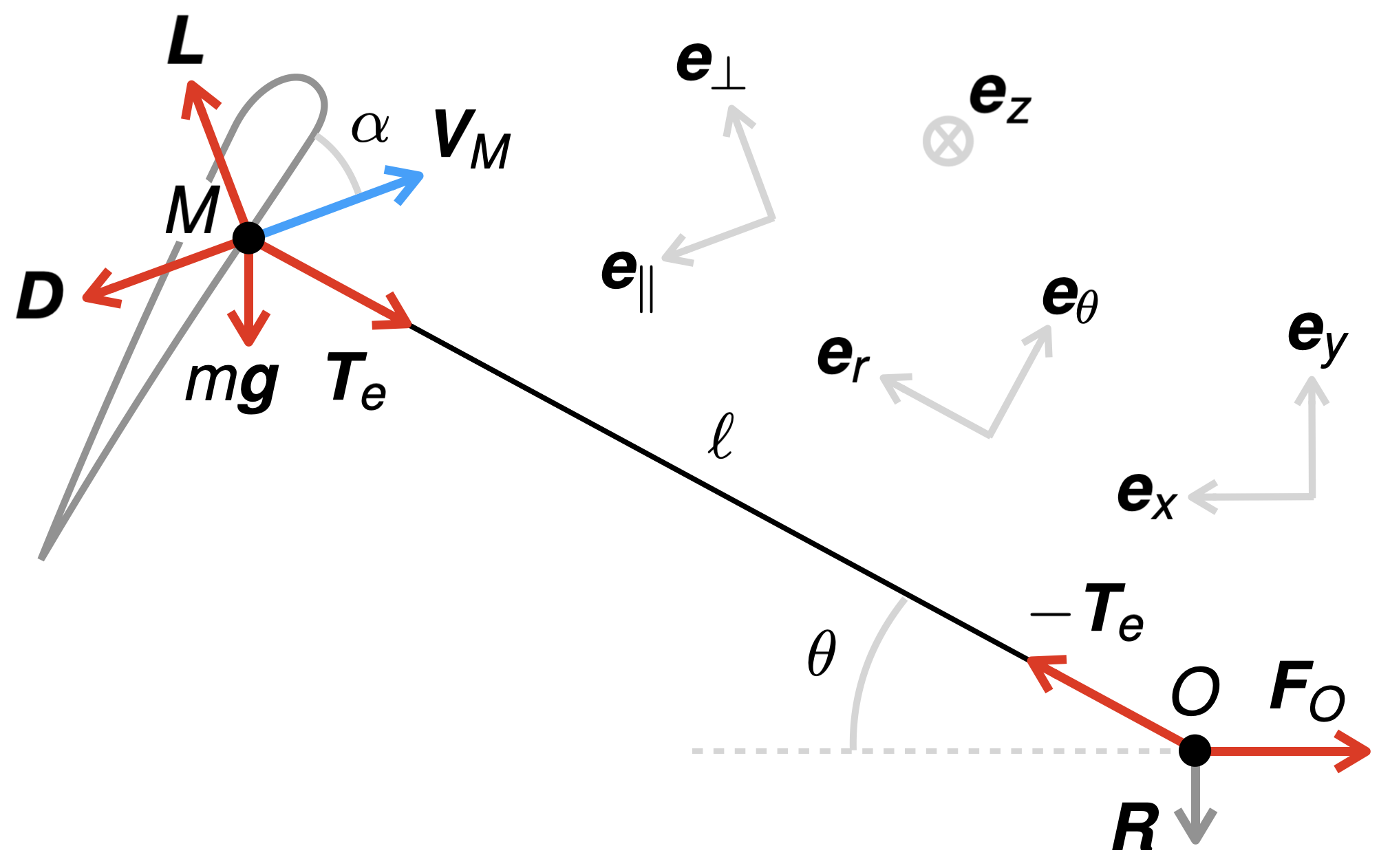}
    \caption{Schematics supporting the theoretical model (see Sec.~\ref{sec:theory}).}
    \label{fig:schem}
\end{figure}

Applying Newton’s second law to point $M$, together with the
angular momentum theorem relative to $O$ yields the following dimensionless equations (see Appendix): 
\begin{eqnarray}
 \dot V_O \cos \theta -\dot \theta^2 +\textstyle \frac1{\cos\theta}&=& \textstyle \frac12\rho S  V_M^2(C_{L} \cos\alpha + C_{D}\sin\alpha)\quad \label{eq:final1}\\
 \ddot \theta +\textstyle  \frac{\tan \theta}{\cos \theta} - \dot \theta^2 \tan \theta &=&\textstyle \frac12\rho S  V_M^2 \big(C_{L}[\sin\alpha +\cos\alpha \tan\theta] \nonumber\\ &&+ C_{D}[\sin\alpha \tan\theta - \cos\alpha]\big),\label{eq:final2}
\end{eqnarray}
with the dimensionless variables $t\rightarrow t/\tau$, $V\rightarrow V\tau/\ell$ with $\tau^2 = m\ell /F_O$, $\rho\rightarrow \rho \ell^3/m$ and $S\rightarrow S/\ell^2$, and where we have neglected gravitational forces.  $F_O$ only appears in the equations through the timescale $\tau$ used to normalize time, thereby explaining the universality of the launch trajectories. For the small-scale glider,  $\tau \approx 0.3[\sqrt{\mathrm{N}}]/\sqrt{F_O}\,$s, while for the real scale UFO wing $\tau \approx 3[\sqrt{\mathrm{N}}]/\sqrt{F_O}\,$s. 
This  timescale enables computation of the typical force that is required to launch in a given amount of time.

Equations~\eqref{eq:final1} and \eqref{eq:final2} constitute a system of two coupled ordinary differential equations, with only $\theta$, $V_O$ and their derivatives as unknowns (see Appendix). They can be jointly solved numerically for given functions $C_L(\alpha)$ and $C_D(\alpha)$. 
However, 
typical lift and drag coefficients are only valid in stationary conditions. As such, they are not expected to stand in the fundamentally unsteady conditions of the launching phase (for one thing, within steady flight theory, the wing should stall in the very beginning of the launch as the angle of incidence $\alpha \to \pi/2$, which is obviously not the case here). Thus Eqs~\eqref{eq:final1} and \eqref{eq:final2} are not expected to provide good quantitative fits of the trajectories. Nonetheless, typical $C_L$ and $C_D$ values found for this type of airfoil (see e.g.~\cite{becker2017experimental}) give good qualitative agreement (see Appendix) and are useful to identify the timescale $\tau$ and prove the experimentally observed universality.

Measuring unsteady lift and drag coefficients would be very interesting, but beyond the scope of this paper. Unsteady $C_L$ and $C_D$  
 are expected to be highly nontrivial as, in the early stages, the wing is not fully inflated, and thus its very shape varies over time (this might explain why the wing doesn't stall).  
We leave these points for future research.

\section{Discussion and Perspectives} \label{sec:conclusion}

Let us first stress the interest of such a study for students. It provides a practical example of the use of Newtonian mechanics and the benefits of nondimensionalization. It also shows how simple experiments and  models can provide an understanding of a new field: the physics of paragliding.

This study revealed a number of other exciting questions.
{In particular, we observed that there is a minimum force $F_O^{\min}$ below which there is no inflation, both in the reduced-scale ($\approx 4$\,N) and real-sized experiments. Consequently, all the results presented here only hold for $F_O> F_O^{\min}$.  Measuring $F_O^{\min}$ is quite difficult because it is very sensitive to variations in the initial conditions, in contrast with the trajectories which appear to be very robust. Characterising $F_O^{\min}$ as function of different parameters, in particular different wings, would certainly be very interesting.}

{Among other interesting questions are} the optimal folding of the wing for a comfortable launch in strong wind conditions, or the differences between regular double-skin and single skin wings during unsteady phases, and the launching phase in particular. {Differences are certainly to be expected regarding the $F_O^{\min}$, which we phenomenologically observed to be larger for regular double skin gliders. For certain gliders, it may even be necessary to  briefly pull on the A risers (front risers acting on the leading edge) to set off the inflation phase.} Further, single-surface gliders are known to launch much faster than their double-skin counterparts. This is often attributed to the fact that they are lighter and thus have less inertial effects, but we believe it can also be related to the fact that the inflation phase is much faster, as argued in Section~\ref{sec:redexp}. 
For regular double skin gliders the  question of the simultaneity for the  air-filling of the chambers and the rising phases, is complicated. Indeed, one expects that the lift coefficient  grows progressively as the chambers fill with air, which explains why the wing starts to take off before the glider is fully inflated. To quantitatively unravel the role of these phases, one could think of combining high speed camera filming of a static inflation experiment  (in which the glider is attached to the ground at its trailing edge), together with an experiment on rigid wings (printed in 3D or cut out in polystyrene), and compute the characteristic times of each isolated phase to disentangle their interactions. One should bare in mind that, while seemingly irrelevant in the present communication, the limitations to scale down the stiffness of the materials in our experiments might have important implications on the characteristics of the inflation phase. These ideas are left for future research.

\section*{Acknowledgements} \label{sec:acknowledgements}
    
We deeply thank Alexandre Darmon for agreeing to serve as test pilot during the real scale experiments and for proofreading the manuscript, as well as Henri Montel (Freedom Parapente, Puy de Dôme) for lending the gliders and several discussions. We also thank Edgar Pereyron for his contribution to improve the algorithm used to solve Eqs.~\eqref{eq:final1} and \eqref{eq:final2}.
The authors have no conflicts to disclose.

\clearpage

 \bibliographystyle{apsrev4-1}

\bibliography{biblio}

\clearpage

\onecolumngrid

\section*{Appendix}

\begin{figure}[h!]
\includegraphics[width=0.45\columnwidth]{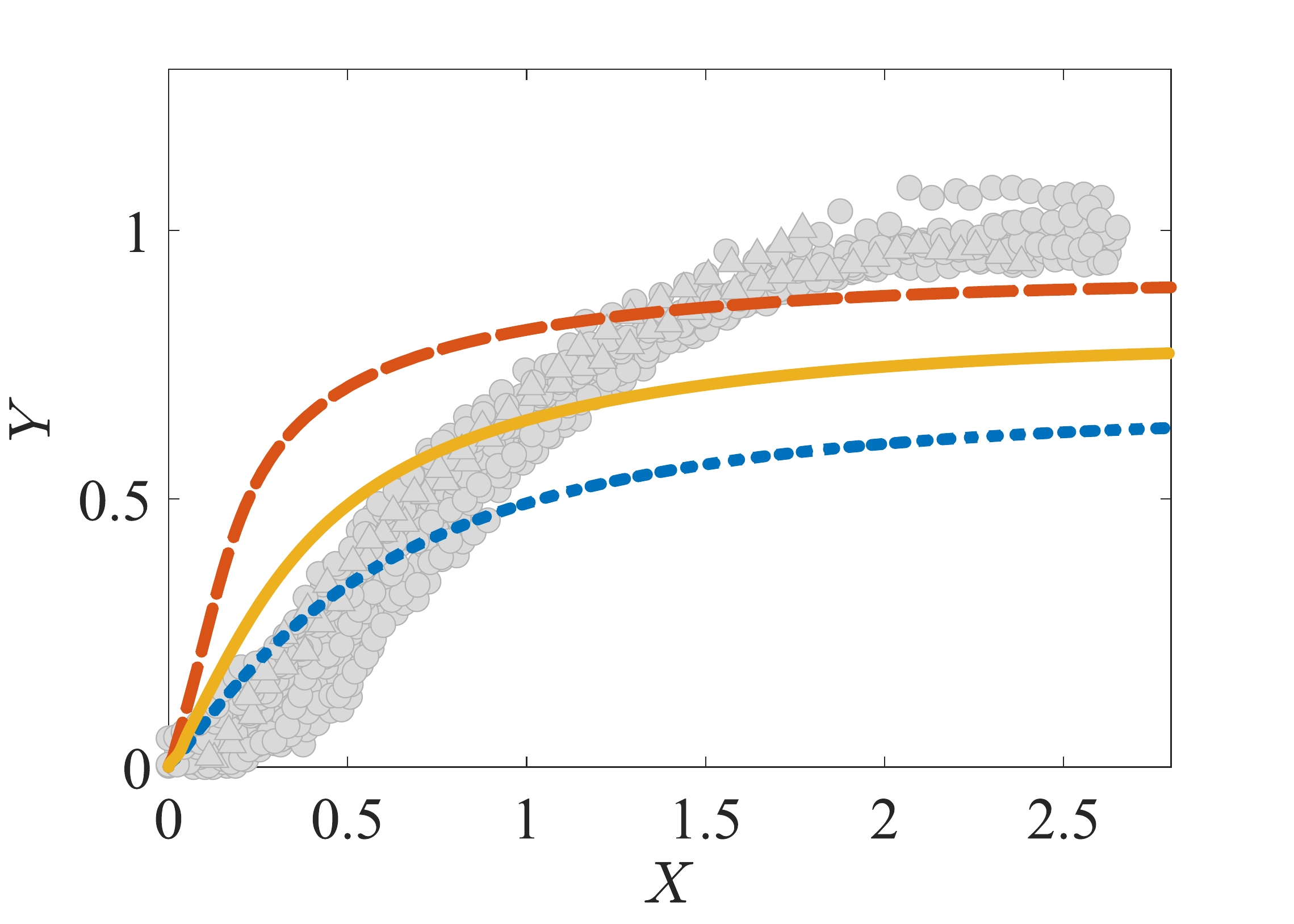}
\captionsetup{justification=raggedright,
singlelinecheck=false
}
    \caption{Numerical solution of Eqs.~\eqref{eq:final1} and \eqref{eq:final2} for $\mu_L=4$, $\mu_D=0.2$ and $A=3$ (solid yellow line), together with the experimental results of Fig.~\ref{fig:masterplot} (light gray markers). The blue and red solutions are obtained by varying the parameters and show that one can either match the initial slope (dotted blue line) by relatively increasing drag (same as solid yellow but with $A=4.5$), or approach the final plateau (dashed red line) by relatively increasing lift (same as solid yellow but with  $\mu_L=8$), but not both.}
    \label{fig:appendix}
\end{figure}

 The angular momentum of the glider relative to $O$ is $m\ell^2 \dot \theta \bs e_z$. The forces acting on $M$ in the non-inertial frame of reference attached to $O$ are the lift and drag forces, as given by Eqs.~\eqref{eq:lift} and \eqref{eq:drag}, the weight $m\bs g$, the tension $- T_e \bs e_r$, and the fictitious force $-m \dot V_O \bs e_x$, where $V_O$ denotes the velocity of point $O$ in the inertial frame of reference. The  angular momentum theorem yieds: 
\begin{eqnarray}
 m\ell \ddot \theta = L \sin\alpha - D \cos \alpha -mg\cos\theta + m\dot V_O\sin \theta, \label{eq:moment1}
\end{eqnarray}
where the right-hand side (upon multiplication by $\ell$) denotes the torque  on $M$ with respect to O along $\bs e_z$. Newton’s second law applied to point $M$ along $\bs e_r$ yields:
\begin{eqnarray}
- m\ell \dot\theta ^2  = L \cos\alpha + D \sin \alpha-mg \sin\theta - T_e -m \dot V_O\cos\theta. \label{eq:pfd1}
\end{eqnarray}
The angle of attack $\alpha$ and the true airspeed $V_M$ are kinematically related to $\theta$ through the velocity composition law: 
\begin{eqnarray}
\cos\alpha = \frac{\ell \dot\theta - V_O\sin\theta}{\sqrt{V_O^2\cos^2\theta+(\ell \dot\theta - V_O\sin\theta)^2} }; \quad \sin\alpha = - \frac{ V_O\cos\theta}{\sqrt{V_O^2\cos^2\theta+(\ell \dot\theta - V_O\sin\theta)^2}};\label{eq:cosalpha}
\end{eqnarray}
\begin{eqnarray}
V_M^2= V_O^2\cos^2\theta + (\ell \dot\theta - V_O\sin\theta)^2.\label{eq:vm}
\end{eqnarray}
Finally, applying Newton's second law to massless point $O$ yields: 
\begin{eqnarray}
T_e = \frac{F_O}{\cos\theta},\label{eq:Te}
\end{eqnarray}
which can be inserted into Eq.~\eqref{eq:pfd1} to eliminate $T_e$. Equations~\eqref{eq:moment1} and \eqref{eq:pfd1} -- combined with Eqs.~\eqref{eq:lift}, \eqref{eq:drag}, \eqref{eq:cosalpha}, \eqref{eq:vm} and \eqref{eq:Te} -- constitute a system of two coupled ordinary differential equations, with only $\theta$, $V_O$ and their derivatives as unknowns.
Isolating $\dot V_0$ in \eqref{eq:pfd1} and inserting it into \eqref{eq:moment1} finally yields:
\begin{eqnarray}
\dot{\tilde{V}}_{O}\cos\theta &=& \dot{\tilde{\theta}}^{2} + \dfrac{1}{2}\tilde{\rho}\tilde{S}\tilde V_{M}^{2}(C_{L}(\alpha) \cos\alpha + C_{D}(\alpha)\sin\alpha) - \tilde g \sin\theta - \frac{1}{\cos\theta}\label{eq:finaltilde1} \\
\ddot{\tilde{\theta}} + \frac{\tan\theta}{\cos\theta} - \dot{\tilde{\theta}}^{2}\tan\theta& =& \frac{1}{2}\tilde{\rho}\tilde{S}\tilde V_{M}^{2}[C_{L}(\alpha)(\sin\alpha + \cos\alpha \tan\theta) + C_{D}(\alpha)(\sin\alpha \tan\theta - \cos\alpha)]\nonumber\\ &&-\tilde g(\cos \theta + \sin \theta\tan\theta)\label{eq:finaltilde2}
\end{eqnarray}
where we have introduced the dimensionless variables $\tilde t= t/\tau$, $\tilde  V= V\tau/\ell$, $\dot{\tilde \theta} = \dot \theta \tau $, $\ddot{\tilde \theta} = \ddot \theta \tau^2 $, $\tilde g = g\tau^2/\ell$, and $\tau^2 = m\ell /F_O$, $\tilde \rho=\rho \ell^3/m$ and $\tilde S = S/\ell^2$. 
Equations~\eqref{eq:finaltilde1} and \eqref{eq:finaltilde2} are equivalent to Eqs.~\eqref{eq:final1} and \eqref{eq:final2} in the main text, where all the $\sim$ have been dropped, and where the gravitational terms have been set to zero, given that their contribution turns out to be negligible (correction of less than $1\%$). A numerical solution can be obtained using a standard RK45 method (explicit Runge-Kutta~\cite{butcher2016numerical}) choosing e.g. $C_L=\mu_L\sin 2\alpha $ and $C_D =\mu_D \sin^2\alpha + A$ where $A$ is a constant  (see~\cite{becker2017experimental}). Figure~\ref{fig:appendix} displays numerical results for different values of $\mu_D$ and $A$. Again,  we do not expect quantitative agreement with the experimental trajectories for the reasons presented in Sec.~\ref{sec:theory}. The theory is only intended to account for the universality, through the identification of the typical timescale~$\tau$.

\end{document}